\documentclass[sigconf]{acmart}

\AtBeginDocument{%
  }

\setcopyright{acmlicensed}
\copyrightyear{2024}
\acmYear{2024}
\acmDOI{XXXXXXX.XXXXXXX}

\usepackage{amsmath}
\usepackage{booktabs}
\usepackage{xcolor}
\usepackage{multirow}
\usepackage{mathtools}
\usepackage{bm}
\usepackage[normalem]{ulem} 

\usepackage{courier}
\usepackage[utf8]{inputenc}
\usepackage{caption}
\usepackage{nicematrix}
\usepackage{fontawesome5}

\acmConference[Conference acronym 'ICMR 2024']{Make sure to enter the correct conference title from your rights confirmation email}{June 10--13, 2024}{Phuket, Thailand}
\acmISBN{978-1-4503-XXXX-X/18/06}

\copyrightyear{2024}
\acmYear{2024}
\setcopyright{rightsretained}
\acmConference[ICMR '24]{Proceedings of the 2024 International Conference on Multimedia Retrieval}{June 10--14, 2024}{Phuket, Thailand} \acmBooktitle{Proceedings of the 2024 International Conference on Multimedia Retrieval (ICMR '24), June 10--14, 2024, Phuket, Thailand}\acmDOI{10.1145/3652583.3658067}
\acmISBN{979-8-4007-0619-6/24/06}

\begin{document}

\title{Anchor-aware Deep Metric Learning for Audio-visual Retrieval}


\author{Donghuo Zeng}
\authornote{Corresponding authors.}
\email{do-zeng@kddi-research.jp}
\affiliation{%
  \institution{KDDI Research, Inc.}
  \streetaddress{Fujimino, Ohara, 2 Chome−1−15}
  \city{Saitama}
  \country{Japan}
  \postcode{356-0003}
}

\author{Yanan Wang}
\email{wa-yanan@kddi-research.jp}
\affiliation{%
  \institution{KDDI Research, Inc.}
  \streetaddress{Fujimino, Ohara, 2 Chome−1−15}
  \city{Saitama}
  \country{Japan}
  \postcode{356-0003}
}

\author{Kazushi Ikeda}
\email{kz-ikeda@kddi-research.jp}
\affiliation{%
  \institution{KDDI Research, Inc.}
  \streetaddress{Fujimino, Ohara, 2 Chome−1−15}
  \city{Saitama}
  \country{Japan}
  \postcode{356-0003}
}

\author{Yi Yu}
\authornotemark[1]
\email{yiyu@hiroshima-u.ac.jp}
\affiliation{%
 \institution{Hiroshima University}
 \streetaddress{1-4-1 Kagamiyama, Higashi-Hiroshima City}
 \city{Hiroshima}
 \country{Japan}
 \postcode{739-8527}
 }

\renewcommand{\shortauthors}{Donghuo Zeng et al.}





\begin{abstract}
Metric learning minimizes the gap between similar (positive) pairs of data points and increases the separation of dissimilar (negative) pairs, aiming at capturing the underlying data structure and enhancing the performance of tasks like audio-visual cross-modal retrieval (AV-CMR). Recent works employ sampling methods to select impactful data points from the embedding space during training. However, the model training fails to fully explore the space due to the scarcity of training data points, resulting in an incomplete representation of the overall positive and negative distributions. In this paper, we propose an innovative Anchor-aware Deep Metric Learning (AADML) method to address this challenge by uncovering the underlying correlations among existing data points, which enhances the quality of the shared embedding space. Specifically, our method establishes a correlation graph-based manifold structure by considering the dependencies between each sample as the anchor and its semantically similar samples. Through dynamic weighting of the correlations within this underlying manifold structure using an attention-driven mechanism, Anchor Awareness (AA) scores are obtained for each anchor. These AA scores serve as data proxies to compute relative distances in metric learning approaches. Extensive experiments conducted on two audio-visual benchmark datasets demonstrate the effectiveness of our proposed AADML method, significantly surpassing state-of-the-art models. Furthermore, we investigate the integration of AA proxies with various metric learning methods, further highlighting the efficacy of our approach. 
\end{abstract}

\maketitle
\section{Introduction}  \label{intro}
Metric learning endeavors to construct an embedded representation of the data, ensuring that similar pairs remain proximate while dissimilar pairs are distanced from each other within the embedding space~\cite{lowe1995similarity, xing2002distance, kulis2013metric, weinberger2005distance}. Metric learning serves as a cornerstone in various domains, such as image retrieval and clustering~\cite{lee2008rank, xing2002distance}, cross-modal learning~\cite{mignon2012cmml}, person re-identification~\cite{hermans2017defense}. 
In practice, as the objective of metric learning methods is to establish an embedding space where similar samples are drawn closer together while dissimilar samples are distinctly separated, it has been applied in AV-CMR task~\cite{zeng2018audio}, by refining the similarity among data samples to enhance the efficacy of retrieval across different modalities.

\begin{figure}[t]
\centering
\includegraphics[width=0.8\columnwidth]{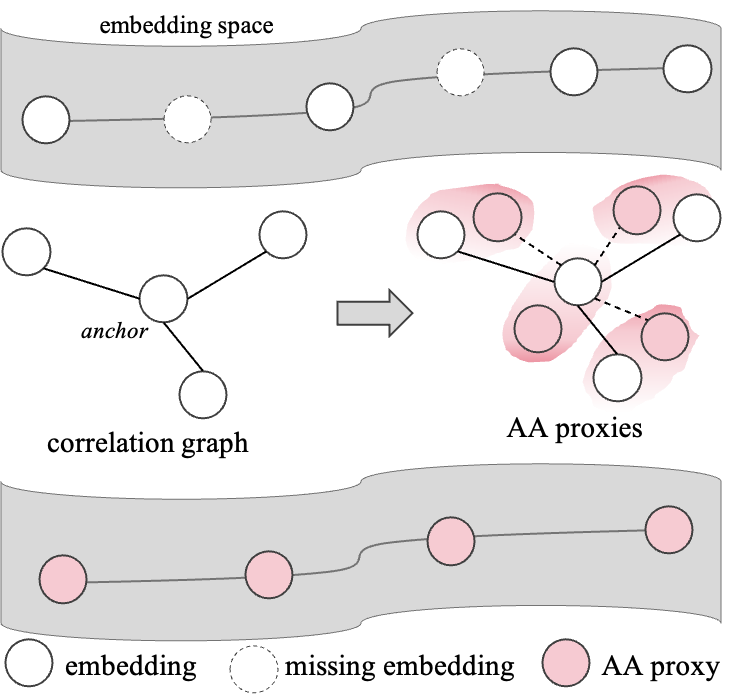} 
\caption{This diagram illustrates the role of the anchor-aware (AA) proxy in deep metric learning. Missing embeddings due to a lack of training data points leads to suboptimal learning of the embedding space. We introduce an AA proxy derived from the correlation graph for each embedding, facilitating the migration toward optimal embedding space learning.}
\label{fig:overview}
\end{figure}

Recent works~\cite{kaya2019deep,kulis2013metric} on metric learning always incorporate deep learning approaches to leverage their capability of learning robust feature representations. As such deep learning-based metric learning approaches~\cite{kaya2019deep} often require large amounts of data to proficiently acquire intricate feature representations, the imbalanced distribution of similar and dissimilar data pairs in the massive training dataset leads to suboptimal results of metric learning. 
Existing works~\cite{zheng2019hardness,ko2020embedding, li2022neighborhood,venkataramanan2021takes} aim to generate synthetic hard negative samples where dissimilar data points are mistakenly identified, by employing adversarial learning~\cite{zheng2019hardness, li2022neighborhood} and linear interpolation~\cite{ko2020embedding} or non-linear interpolation~\cite{venkataramanan2021takes}, which captures the potential of numerous easy negatives. 
However, these works face challenges in controlling the number of insertion points to obtain optimal models and often overlook the need to identify robust anchors to fully capture the positive and negative distribution to refine the embedding space. Overall, the scarcity of training data points will raise the problem of missing embeddings in existing methods. This problem leads to suboptimal learning of the embedding space and compromises the quality of sample representation, including positive and negative samples, consequently impairing the performance of deep metric learning.
 
To address the issue, we propose an innovative Anchor-aware Deep Metric Learning (AADML) method to uncover the intrinsic correlations among existing data points and compensate for the incomplete learning of the embedding space due to insufficient data points, as illustrated in Fig.~\ref{fig:overview}. Our approach comprises three fundamental components: (1) as the manifold signifies the foundational structure and relationships within the data, we construct a correlation graph within each modality to facilitate the adept capture of the innate manifold structure. (2) To effectively capture correlations among similar samples and comprehensively consider each sample's contribution to the embedding space learning process, we leverage manifold pairs from the correlation graph across modalities using an attention-driven mechanism. For example, if we select an audio sample as an anchor point and identity the $K$ nearest audio samples relative to the anchor from the correlation graph, we create $k$ tuples as inputs of an attention-driven model to generate an anchor-aware (AA) proxy for the anchor. (3) Leveraging AA proxies due to their ability to directly capture nuanced dependencies among similar pairs while indirectly mitigating the interference of dissimilar pairs, we employ these AA scores as sample proxies to calculate relative distances within a metric learning framework.

To validate the effectiveness of the proposed AADML approach, extensive experiments are conducted on two cross-validated audio-visual benchmark datasets. These datasets encompass annotation labels present in both audio and visual modalities and offer a comprehensive evaluation setup. Experimental results exhibit the superiority of our anchor-aware approach, achieving substantial performance enhancements, and outperforming state-of-the-art models in AV-CMR tasks by 3.0\% on large dataset VEGAS and by 45.6\% on small dataset AVE in terms of mean average precision (MAP), respectively. Additionally, we demonstrate the applicability of the AA proxy across various metric learning methods, resulting in performance surpassing existing state-of-the-art losses in AV-CMR tasks.

The remainder of this paper unfolds as follows: Section 2 provides an overview of related works. Section 3 elaborates on the proposed Anchor-Aware Deep Metric Learning (AADML) approach and Section 4 outlines the experimental setup and results. Lastly, in Section 6, we summarize our work and discuss potential future directions in the AV-CMR task.

\section{Related Work} \label{relatedwork}
\subsection{Audio-visual Cross-modal Retrieval}
Audio-visual cross-modal retrieval is a challenging task that aims to retrieve relevant content across different modalities such as audio and visual modalities\cite{suris2018cross, zeng2018audio,zeng2022complete}.
Canonical Correlation Analysis (CCA)~\cite{hardoon2004canonical} is a classical linear technique used to find correlated patterns between two sets of variables. In the context of AV-CMR, CCA-based methods have been extensively studied to learn joint representations that capture the correlations between audio and visual data. Representative approaches include traditional CCA~\cite{hardoon2004canonical}, K-CCA~\cite{akaho2006kernel}, Cluster-CCA~\cite{rasiwasia2014cluster}, Deep CCA (DCCA)~\cite{andrew2013deep}, C-DCCA~\cite{yu2018category}, Triplet with Cluster-CCA (TNN-C-CCA)~\cite{zeng2020deep}, and VAE-CCA~\cite{zhang2022}. While CCA-based methods offer interpretable embeddings, they struggle to capture complex and non-linear relationships present in real-world audio-visual data.

Deep learning is a powerful approach for representation learning in AV-CMR, which employs deep neural networks to learn joint embeddings by minimizing the distance between corresponding samples. Notably, the CLIP~\cite{pmlr_v139_radford21a}, ACMR~\cite{wang2017adversarial}, DSCMR~\cite{zhen2019deep}, BiC-Net~\cite{hou2021bicnet}, DCIL~\cite{zheng2020dual}, CCTL~\cite{zeng2022complete}, EICS~\cite{zeng2023learning}, and MSNSCA \cite{zhang2023multi}. These models have exhibited exceptional performance in cross-modal retrieval between audio-visual~\cite{zeng2022complete, zeng2023learning, zhang2023multi}, positioning them as valuable contributions to the field of representation learning. While prior efforts emphasize training intricate neural networks to acquire representations for AV-CMR, our approach introduces a novel direction by focusing on advanced metric learning techniques.

\subsection{Metric Learning for Retrieval}
Metric learning is of paramount importance in retrieval tasks, particularly when tailored to the specific requirements of AV-CMR. The objective is to develop a suitable distance or similarity metric that preserves the underlying structure of data across different modalities to enhance retrieval accuracy. 

Traditional metric learning methods such as Mahalanobis distances~\cite{de2000mahalanobis}, focus on preserving the intra-class similarity while increasing the inter-class distance. However, they may struggle with capturing non-linear relationships within complex audio-visual data. The emergence of deep metric learning has gained significant traction across various retrieval tasks, including AV-CMR. Prominent techniques, such as contrastive loss~\cite{hadsell2006dimensionality} and triplet loss~\cite{hermans2017defense}, aim to pull similar samples closer together and push dissimilar samples further apart in the learned embedding space. These methods excel in generating discriminative embeddings by analyzing relative distances between samples. Additionally, an array of deep metric learning variations~\cite{hoffer2015deep, kaya2019deep, ko2020embedding, li2022neighborhood, venkataramanan2021takes, zheng2019hardness} has been introduced. These novel deep metric learning methods offer the potential to further enhance AV-CMR, For instance, the CCTL~\cite{zeng2022complete} model adapts triplet loss to audio-visual cross-modal data. However, it has limitations in terms of model training difficulty and sensitivity to data scarcity situations.  

The works~\cite{ermolov2022hyperbolic, vinh2020hyperml, kim2023hier} have proposed advanced metric learning in some clues. Notably, approaches such as harnessing hyperbolic geometry to capture the intricate correlations present in natural data have emerged~\cite{ermolov2022hyperbolic, vinh2020hyperml, kim2023hier}, which exists its potential complexity and computational overhead. Similar problems will happen in the generator model involved in data augmentation~\cite{zheng2019hardness, ko2020embedding, liu2022densely} methods. These methods are an effective strategy to address data scarcity by generating synthetic samples through various transformations. However, the effectiveness of data augmentation is sensitive to the quality of synthetic points and the specific generation methods employed, raising concerns about model reliability.


Instead of relying on data augmentation or mining effective data settings to train by complex selection methods, our proposed approach efficiently computes global attention-driven dependencies between an anchor and other similar samples in parallel to dynamically weigh the correlations of their manifold structure, the correlations scores as proxies for data samples will improve the performance of AV-CMR while using such proxies for the deep metric learning.

\begin{figure*}[t]
\centering
\includegraphics[width=\textwidth]{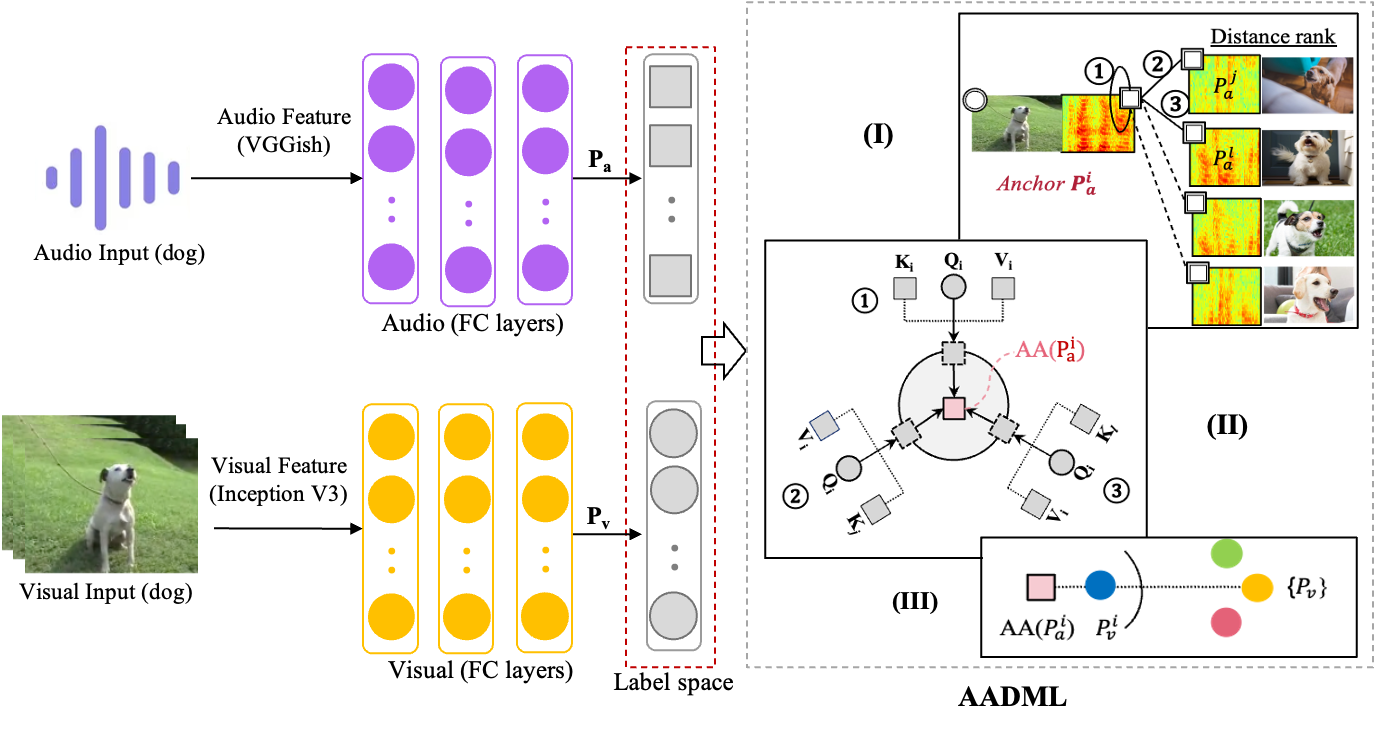} 
\caption{The framework of our proposed model. The audio and visual features extracted by pre-trained models VGGish and Inception V3, respectively, are projected into label space as predicted label embeddings. AADML approach operates within the label space and comprises three distinct components: (I) Choosing an audio sample $P_{a}^{i}$ as the anchor, we traverse the correlation graph to discern the $k$ (k=3) nearest audio samples ($P_{a}^{k}$ vs. $P_{a}^{l}$) relative to the anchor, thus forming three manifold pairs as key-value pairs for (II), to compute the attention score $A(\cdot)$ while the anchor as query ($Q_{i}$) with each pair: $P_{a}^{\in \{i, j, k\}}$ as key ($K_{\in \{i, j, k\}}$), $P_{a}^{i}$ as value ($V_{i}$). The anchor-aware $AA(\cdot)$ score (pink box) is then obtained as the average of this $ A (\cdot)$ across the three key-value pairs. (III) This score is subsequently utilized as an anchor proxy for foundational metric learning methods like contrastive and triplet loss.}
\label{fig:framework}
\end{figure*}

\section{Problem Formulation}
In audio and visual cross-modal retrieval tasks, the primary challenge arises from the intricate features of audio and visual data, coupled with their different feature distributions. This disparity makes direct comparisons using basic metrics such as distance or similarity metrics unavailable. Hence, there arises a need to devise cross-modal retrieval methods. These methods aim to map audio and visual features into a shared embedding space, facilitating the generation of new features for each modality. These features can then be utilized for comparison computations directly.

Assume an audio-visual dataset comprising $n$ videos, symbolized as $\Sigma=\{\sigma_{i}\}_{i=0}^{n-1}$, where each video consists an audio and visual pair, represented as $\sigma_{i}=(a_{i}, v_{i})$. Here, $a_{i}\in R^{128}$ denotes audio features extracted by VGGish pre-trained model\footnote{https://github.com/tensorflow/models/blob/master/research/audioset/vggish/README.md}, and $v_{i}\in R^{1024}$ signifies visual features extracted by Inception V3 pre-trained model\footnote{https://huggingface.co/docs/timm/en/models/inception-v3}. Each $\sigma_{i}$ instance is associated with a semantic vector label $Y_{i}=[y_{i}^{0}, y_{i}^{1}, ..., y_{i}^{c-1}]$, where $c$ represents the number of semantic categories. In this context, $y_{i}^{j}$ is a binary value; it equals 1 if the sample belongs to the $j$-th category ($j=0, 1, 2, ..., c-1$), and 0 otherwise.

Our objective is to feed the $a_{i}$ and $v_{i}$ pair into a shared label space, where the feature dimension aligns with the number of pre-defined labels. To enhance the learning of shared features, we introduce a novel metric learning approach aligned with triplet loss. This method creates a transformative proxy for each anchor. For instance, if we designate an audio sample $a_{i}$ as an anchor, its transformative proxy is denoted as $AA(a_{i})$. This proxy captures intricate dependencies and generates associations through attention-driven mechanisms. Empowered by $AA(a_{i})$, samples belonging to the same class collaborate to recalibrate and refine their embeddings. This collective effort culminates in the creation of a more nuanced, comprehensive, and accurate semantic embedding space. Consequently, it enhances the performance of fundamental metric learning techniques.

\section{Approach} \label{mymodel}
\label{approachsection}
In this section, we comprehensively present an Anchor-Aware Deep Metric Learning (AADML) model, which is a cohesive framework that combines: \ref{manifold}. correlation graph-based manifold structure, \ref{proxy}. anchor-aware (AA) proxy, and \ref{leverage}. leveraging the AA proxy in metric learning to enhance the accuracy of AV-CMR, as seen in Fig.~\ref{fig:framework}.

\subsection{Correlation Graph-based Manifold Structure}
\label{manifold}
The utilization of manifold structure is crucial for facilitating our Anchor-Aware mechanism to accurately capture dependencies of similar pairs. We employ an approach based on correlation graphs to capture the underlying manifold structure inherent in data samples. The objective of this approach is to ensure that the similarity distance between data samples within the same manifold remains small, thereby promoting retrieval accuracy.

We commence by constructing undirected graphs, $Graph = (V_{m}, S_{m})$, $m$ individually for the audio or visual modality. Here, $V_{m}$ can be audio set $\{a_{0}, a_{1}, ..., a_{n-1}\}$ or visual set $\{v_{0}, v_{1}, ..., v_{n-1}\}$, which represents the vertices, and $S_{m}$ constitutes the cosine similarity matrix of $m$ modality. The matrix $S_{m}$ is meticulously defined as follows.

\begin{equation}
  S_{m}(p, q) =
    \begin{cases}
      1 & \text{if}~x_{p} \in k$-$NN(x_{q})~\text{and}~(p, q) \in {C}_{i} \\
      0 & \text{Otherwise} 
    \end{cases}   
\end{equation}

where $p$ and $q$ are pointers ranging from 0 to $n$-$1$, $x\in \{a, v\}$, $x_{q}$ and $x_{p}$ indicate $q$-$th$ and $p$-$th$ of $x$, respectively. $(p, q)\in {C}_{i}$ represents $x_{p}$ and $x_{q}$ belong to the same category ${C}_{i}$, $i=0, 1, ..., c-1$. The $k$-$NN(x_{q})$ is the k-nearest neighbor of $x_{q}$ within the training set. The k-nearest neighbors algorithm consists of three main steps for identifying the nearest neighbors to a given reference vector~$x_{q}$. 

Firstly, we compute the cosine similarity between the $x_{q}$ for one modality and each vector $x_{p}$ from another modality in the dataset. The cosine similarity between two vectors $x_{q}$ and $x_{p}$ is defined in Eq.~\ref{eq:cosine}:
\begin{equation}
\text{Similarity}(x_q, x_p) = \frac{x_q \cdot x_p}{\|x_q\| \cdot \|x_p\|}
\label{eq:cosine}
\end{equation}

Secondly, after computing the cosine similarity for each pair $(x_{q}, x_{p})$, we rank the computed similarities for all pairs by decreasing order as a rank list, which is computed by Eq.~\ref{eq:rank}.
\begin{equation}
\text{rank}(x_q, x_p) = \text{argsort}_{p=0}^{n-1}(Similarity(x_q, x_p))
~\label{eq:rank}
\end{equation}
where given a query $x_q$ and fixed the indicator $q$, we rank the similarity between $x_q$ and all the $x_p$, $p=0, 1, .., n-1$.

Thirdly, we select the top $k$ similarities from the ranked list $rank(x_q, x_p)$ to obtain the $k$-nearest neighbors. These selected neighbors, denoted as the set $k$-$NN(x_{q})$, can be represented in Eq.~\ref{eq:select}

\begin{equation}
\text{k-NN}(\mathbf{x}_q) = \{\mathbf{x}_{p1}, \mathbf{x}_{p2}, \ldots, \mathbf{x}_{pk}\}
\label{eq:select}
\end{equation}
where $p1, p2, ..,pk$ are the top $k$ indices of rank list. 

Notably, pairwise information inherently exists in cross-modal data, if the corresponding audio $a_k$ is within the same manifold as audio query $q_j$, the paired visual $v_k$ is also within the same manifold as $q_j$, and vice versa, seen in component (I) of the Fig.~\ref{fig:framework}. With this definition, we aim to leverage the underlying data manifold of different modalities to construct the anchor-aware proxy.

\subsection{Anchor-Aware (AA) Proxy}
\label{proxy}
After obtaining the generated manifold pairs, the method proceeds to predict the relevance score of the anchor by leveraging its corresponding manifold pairs using the attention model. The relevance score is represented as an AA proxy for each anchor. 

\subsubsection{Correlation Graph for AA Proxy}
In this context, we select a sample from one modality as the anchor and then identity the $k$ nearest audio samples relative to the anchor from the \(k\)-NN\((\mathbf{x})\)(Eq.~\ref{eq:select}) in the correlation graph. This process forms three manifold pairs as key-value pairs for the attention-driven mechanism, which computes the AA proxy score for the anchor. 

For instance, suppose we select the audio sample \(P_a^{i}\) as the designated anchor point and set the hyperparameter $k$ as 3, we derive three manifold pairs as key-value pairs: $(P_a^{i}, P_a^{i})$, $(P_a^{j}, P_a^{i})$, and $(P_a^{l}, P_a^{i})$. These pairs are then used as inputs for the attention-driven mechanism to compute the AA proxy score for the anchor \(P_a^{i}\). The resulting tuples are expressed as $(Q, K, V) = \{ (P_v^{i}, P_a^{i}, P_a^{i}), (P_v^{i}, P_a^{j}, P_a^{i}), \\ (P_v^{i}, P_a^{l}, P_a^{i}) \}$.

\subsubsection{Anchor-Aware (AA) Proxy Computation Method}
Drawing inspiration from established concepts in scaled dot-product attention and multi-head attention models, our goal is to capture intricate dependencies within modalities and interdependencies across modalities, while simultaneously extracting adaptive weights for each data point. To achieve this, we utilize scaled dot-product attention to compute attention scores. Then, we employ a multi-head self-attention mechanism in a parallel manner. 

Firstly, we aim to compute the attention score $A(Q, K, V)$ for each tuple $(Q, K, V)$ by using the scaled dot-product attention mechanism, which is a fundamental aspect of AA proxy computation. The $A(Q, K, V)$ score quantifies the degree of focus to allocate to other samples within the input sequence, defined by Eq.~\ref{scaled}

\begin{equation}
\begin{aligned}
A(Q, K, V)&= \operatorname{softmax}\left(\frac{Q W^Q K^TW^K}{\sqrt{d_k}}\right)V W^V
\end{aligned}
\label{scaled}
\end{equation}
where $W^Q$, $W^K$, and $W^V$ are the parameter matrices of $Q$, $K$, and $V$, respectively.

Secondly, multi-head attention focuses on different parts of the input simultaneously by splitting the input into multiple "heads" and computing separate attention scores by projecting the queries, keys, and values $h$ times.
\begin{equation}
\hat{A}(Q, K, V) = \operatorname{Concat}\left(A_1, A_2, \ldots, A_h\right)W^O \\
\end{equation}
where $h$ is the number of attention heads, $W^O$ represents the output projection matrix, which is applied to the concatenated attention heads to compute the final output of the multi-head attention layer.

Finally, we compute the $AA(\cdot)$ proxy, using $a_i$ as the input example.
\begin{equation}
AA(a_{i}) = \frac{1}{k} \sum_{\theta=1}^{k} \hat{A}(v_{i}, a_{\theta}, a_{i}) \\
\end{equation}
where $k$ is the hyperparameter of Eq.~\ref{eq:select}.

The essence of AA proxy lies in establishing global dependencies between anchors (queries) and their semantic counterparts across both intra- and inter-modal relationships. Computed in parallel, these relationships yield a dynamic AA proxy score that encapsulates the degree of association between anchor points and analogous samples. 

\subsection{Leveraging the AA Proxy in Metric Learning}
\label{leverage}
We explore the pragmatic applications of the AA proxy within the realm of metric learning. With $AA(\cdot)$ positioned as an innovative anchor proxy, we redefine conventional metric learning techniques to align with it. Specifically, we demonstrate the seamless integration of $AA(\cdot)$ in computing relative distances for both triplet loss~\cite{hermans2017defense} and contrastive loss~\cite{hadsell2006dimensionality}.

The AA proxy effectively replaces the conventional anchor, while the positive and negative components remain unaltered. This reformulated paradigm infuses the metric learning process with heightened sensitivity to the intricate dependencies captured by $AA(\cdot)$, and harnesses the inherent parallelism within the $AA(\cdot)$, thereby yielding more accurate and robust embeddings. The $AA(\cdot)$, acting as a conduit for similarity measurement, are adeptly utilized as proxies for computing relative distances. For example, Triplet loss with AA proxy can be defined as:
\begin{equation}
\begin{split} 
\mathcal{L}_{\text{AA+triplet}} &= \sum_{i=1}^{N} [ \left\| \mathbf{AA}(a_i) - \mathbf{AA}(p_i) \right\|_2^2 \\
&- \left\| \mathbf{AA}(a_i) - \mathbf{AA}(n_i) \right\|_2^2 + \alpha ]_+
\end{split}
\end{equation}
where $N$ is the number of sample with one batch, ${p_{i}}$ and ${n_{i}}$ are the positive and negative sets of original anchor sample $a_{i}$. The hyperparameter $\alpha$ serves as the margin in triplet loss, defining a threshold that regulates the degree of dissimilarity permissible between the anchor-positive and anchor-negative pairs. The $\rVert \cdot \rVert_2$ represents the Frobenius norm.  The formula for contrastive loss can be expressed as:
\begin{equation}
\begin{split} 
L_{\text{AA+contrastive}} &= (1 - y) \frac{1}{2} D^2 + y \frac{1}{2} \max(0, m - D)^2 \\
D &= \lVert \mathbf{AA}(x_i) - \mathbf{AA}(x_j)  \rVert_2^2
\end{split}
\end{equation}
where $y$ indicate the given two data points $x_i$ and $x_j$ is similar (y=1) or dissimilar (y=0). $m$ is the margin hyperparameter that defines the minimum distance between dissimilar pairs.

Our final objective loss of AA in metric learning, such as triplet loss (TL) or contrastive loss (CL), is defined as follows:
\begin{equation}
\begin{split} 
   Loss &= Loss_{label} + Loss_{AA+TL/CL} \\
   Loss_{label} &= \frac{1}{n}||f_{a}(a_{i})-Y(a_{i})||_{F} + \frac{1}{n}||f_{v}(v_{i})-Y(v_{i})||_{F}
    \label{equ:overall_formula}
\end{split}
\end{equation}
where $||\cdot||_{F}$ signifies the Frobenius norm, $f(x)$ represents the projected feature in the shared label space and $Y(\cdot)$ denotes the label representations. In the end, the optimization of the final objective loss is carried out using the stochastic gradient descent (SGD) algorithm.
\section{Experiment}\label{experiment}
\begin{table*}[t]
\centering

\begin{NiceTabular}{l|ccc|ccc}
\toprule
\Block{2-1}{\textbf{Models}} &  \Block{1-3}{\textbf{VEGAS Dataset}} &&  &\Block{1-3}{\textbf{AVE Dataset}} \\ \cline{2-7}
                       & A$\rightarrow$V & V$\rightarrow$A  & Avg.
                       & A$\rightarrow$V & V$\rightarrow$A  & Avg. \\ \hline
    Random case
    & 0.110 & 0.109 & 0.109     
    & 0.127 & 0.124 & 0.126\\
    CCA~\cite{hardoon2004canonical} 
    & 0.332 & 0.327  & 0.330    
    & 0.190 & 0.189 & 0.190 \\
    KCCA~\cite{akaho2006kernel} 
    & 0.288 & 0.273  & 0.281    
    & 0.133 & 0.135 & 0.134 \\
    DCCA~\cite{andrew2013deep}  
    & 0.478 & 0.457  & 0.468   
    & 0.221 & 0.223 & 0.222 \\
    C-CCA~\cite{rasiwasia2014cluster}  
    & 0.711 & 0.704  & 0.708    
    & 0.228 & 0.226 & 0.227\\
    C-DCCA~\cite{yu2018category}
    & 0.722 & 0.716  & 0.719    
    & 0.230 & 0.227 & 0.229\\
    TNN-C-CCA~\cite{zeng2020deep} 
    & 0.751 & 0.738 & 0.745      
    & 0.253 & 0.258 & 0.256\\ 
    VAE-CCA~\cite{zhang2022} &0.821 &0.824 &0.822
    & 0.328 &0.302 &0.315\\ \hline
    AGAH~\cite{GuGGLXW19}   
    & 0.578 & 0.568 & 0.573      
    & 0.200 & 0.196 & 0.198\\
    ACMR~\cite{wang2017adversarial}   
    & 0.465 & 0.442 & 0.454      
    &  0.162   & 0.159   & 0.161\\
    DSCMR~\cite{zhen2019deep}  
    & 0.732 & 0.721 & 0.727      
    & 0.314 & 0.256 & 0.285\\
    CLIP~\cite{pmlr_v139_radford21a}
    & 0.473 & 0.617 & 0.545      
    & 0.129 & 0.161 & 0.145\\
    BiC-Net~\cite{hou2021bicnet}
    & 0.680 & 0.653 & 0.667      
    & 0.188 & 0.187 & 0.188\\
    DCIL~\cite{zheng2020dual}
    & 0.726 & 0.722 & 0.724      
    & 0.244 & 0.213 & 0.228\\
    CCTL~\cite{zeng2022complete}  & 0.766 & 0.765 & 0.766 & 0.328 & 0.267 & 0.298  \\ 
    VideoAdviser~\cite{wang2023videoadviser}  &0.825 & 0.819 & 0.822 &- &- &-\\
    EICS~\cite{zeng2023learning}  & 0.797 & 0.779 & 0.788 &0.337 & 0.279 & 0.308 \\
    TLCA~\cite{zeng2023two} &0.822 &0.838 &0.830 &\uline{0.410} &\uline{0.451} &\uline{0.431} \\
 MSNSCA~\cite{zhang2023multi}&\uline{0.866}&\uline{0.865}&\uline{0.866} &0.323 &0.343 &0.333 \\
     \hline
    \textbf{AADML (Ours)}
    &\textbf{0.901} &\textbf{0.891} &\textbf{0.896}
    &\textbf{0.890} &\textbf{0.883} &\textbf{0.887}\\
    \bottomrule
\end{NiceTabular}
\caption{Model comparison: MAP scores of our approach versus state-of-the-art methods. The best MAP scores are presented in bold, and the second-best MAP scores are indicated with underlining.}
\label{table:comparison}
\end{table*}

In this section, we conducted experiments to evaluate our proposed AADML model in AV-CMR tasks, we compared our method with the existing state-of-the-art methods. In order to identify whether our method can be useful for the improvement of exiting deep metric learning methods, we combine our method with them to further show the power of our method. We also perform some ablation studies to analyze the effectiveness of our method.

\subsection{Dataset and Metrics}
Our model achieved success in the AV-CMR task based on the assumption that audio and visual modalities share identical semantic information. As a result, we select video datasets containing audio-visual tracks and ensure that both tracks are labeled identically on the time series. We select two special video datasets: VEGAS~\cite{zhou2018visual} and AVE~\cite{tian2018audio}. Both datasets underwent a thorough process of label double-checking, ensuring uniform labeling across all frames in both modalities. The VEGAS dataset consists of 28,103 videos annotated with 10 labels, while the AVE dataset includes 1,955 videos annotated with 15 labels. We adhere to the same data partitioning strategy for training and testing sets, as well as the identical approach for feature extraction as described in the referenced work~\cite{zeng2020deep}. 

For model evaluation metrics, we adopt the same measures as those works~\cite{zeng2023learning, zeng2022complete}: mean average precision (MAP) and Precision-scope@K. The MAP functions as an evaluative metric to gauge the effectiveness of models in AV-CMR. This entails calculating the average precision (AP) for each individual query and subsequently deriving the mean value from these AP scores, providing a comprehensive measure of overall performance. The Precision-scope@K quantifies the proportion of relevant items retrieved out of the total relevant items, up to a specified rank "K" in the ranked list. The model performs better as the values of both metrics increase.

\subsection{Implementation Details}
In the subnetwork of our model, we develop three fully connected (FC) layers and one predicted layer as a decoder for both audio and visual inputs, each fully connected layer was configured with 1024 hidden units, and set dropout as 0.1. To project audio and visual features into a shared label space, we ensure that the feature dimension of the projected features in the label space matches the number of labels. Specifically, we maintain both the predicted and pre-defined categories at 10 for VEGAS and 15 for AVE. We achieved optimal results with batch sizes of 400 and 200, respectively, and conducted training over 400 epochs. For the margin in triplet loss and metric learning, the best results were obtained with values of 1.2 and 1.0, respectively.

We implement all the experiments of our methods with the machine learning framework PyTorch~\footnote{https://pytorch.org/}, and based on Ubuntu Linux 22.04.2 with NVIDIA GeForce 3080 (10G). We utilize the Adam optimizer~\cite{kingma2014adam} with its default parameter configuration to train our model, while the learning rate is designated as 0.0001.

\begin{figure}[t]
    \centering
    \centering {\includegraphics[width=4.1cm]{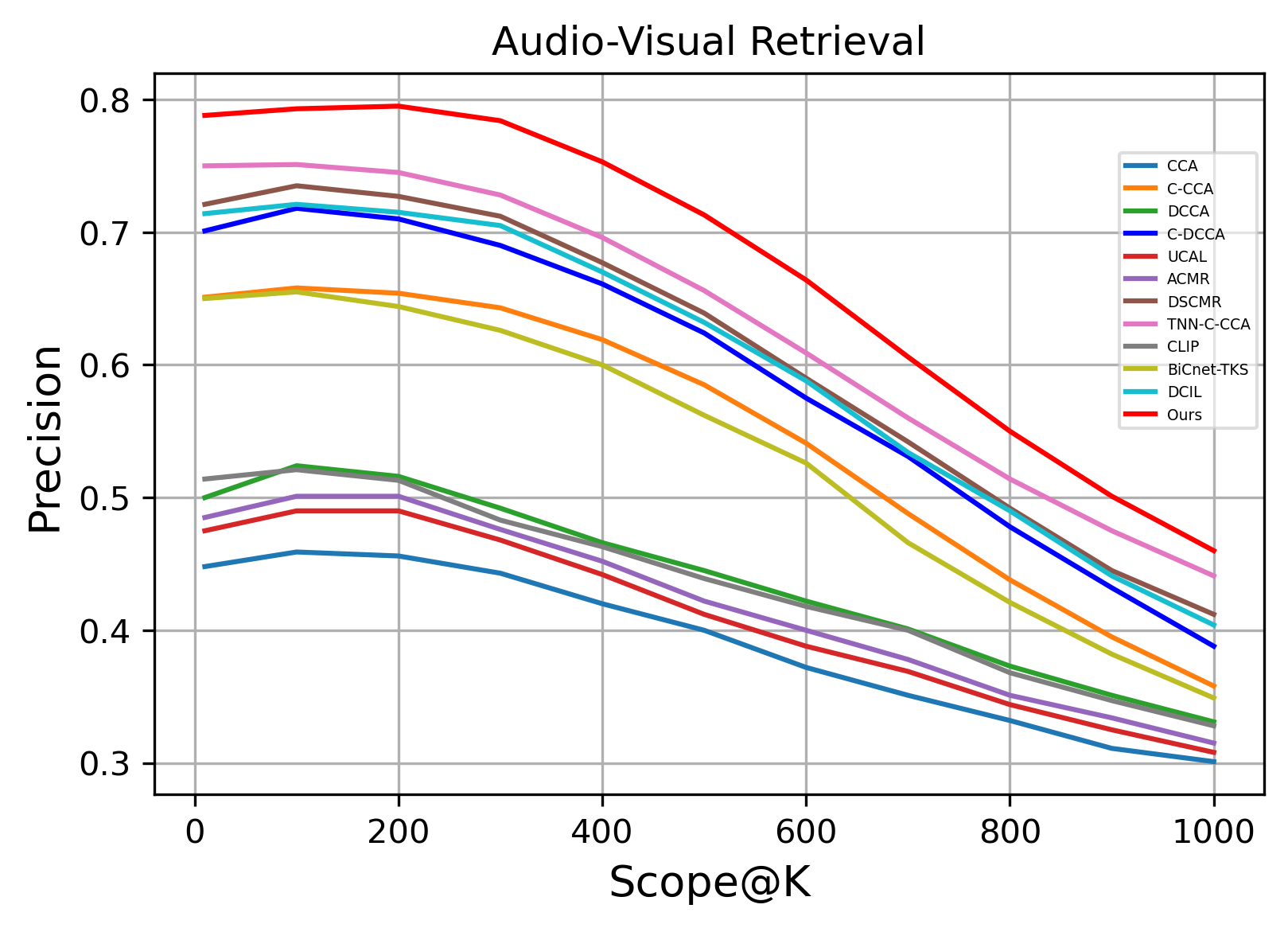}}
    \centering {\includegraphics[width=4.1cm]{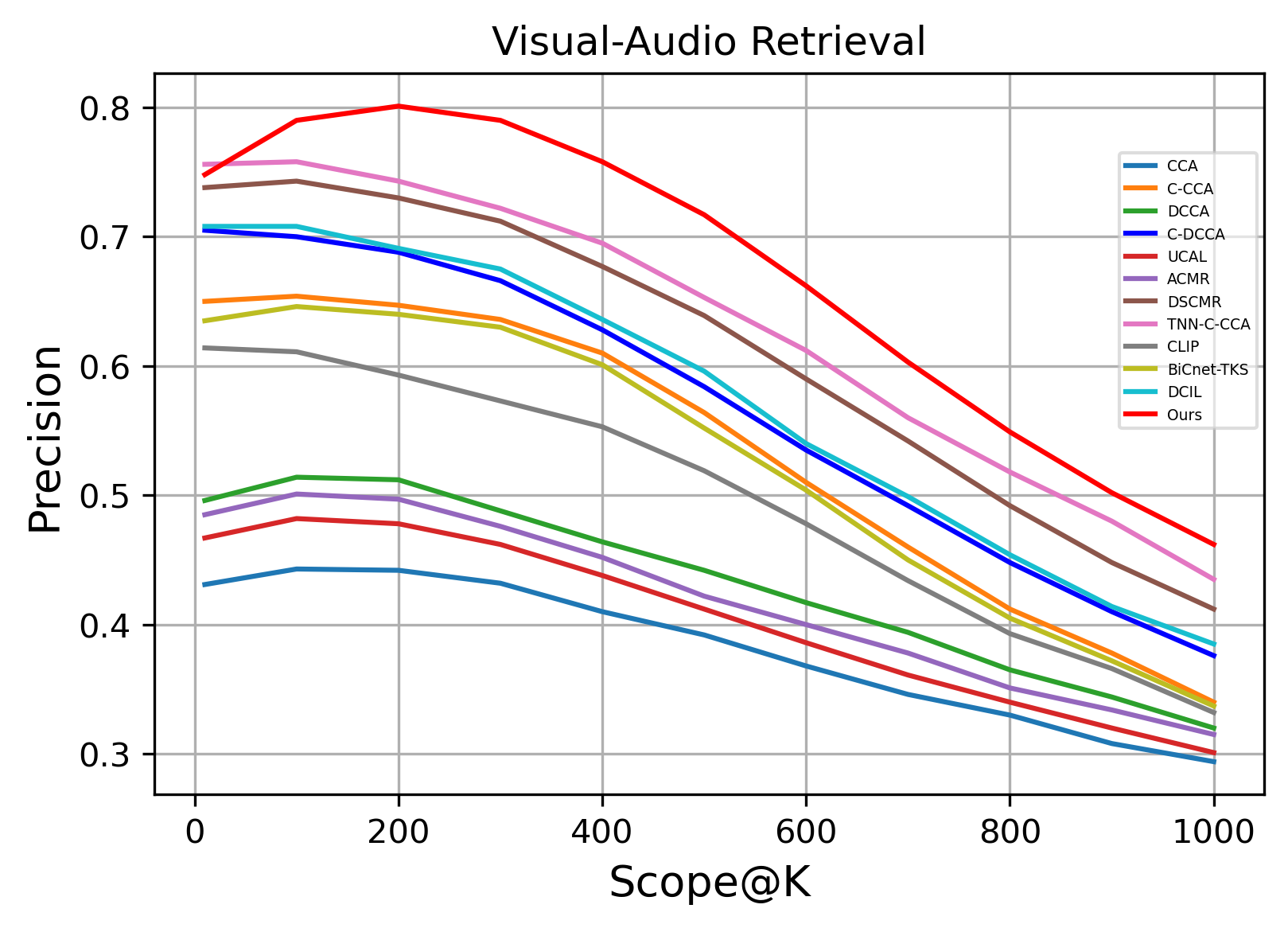}}%
    \caption{Precision-scope@K curves on the VEGAS dataset for $audio\! \rightarrow \! visual$ and $visual\! \rightarrow \! audio$ retrieval experiments, spanning different values of $K$ from 10 to 1000.}%
    \label{fig:PSC}%
\end{figure}
\subsection{Comparison with Existing AV-CMR Approaches} For the purpose of evaluating the efficacy of our proposed method, a comprehensive comparison is conducted against a repertoire of 17 distinct algorithms. This ensemble encompasses seven methods grounded in Canonical Correlation Analysis (CCA), as well as eleven leading-edge techniques rooted in deep learning for cross-modal models. CCA~\cite{hardoon2004canonical} is employed to derive linear transformations for two data sets to optimize their correlation. KCCA~\cite{akaho2006kernel} and DCCA~\cite{andrew2013deep} try to learn non-linear transformations by using the kernel method and deep learning, respectively. C-CCA~\cite{rasiwasia2014cluster} and C-DCCA~\cite{yu2018category} aim to discover transformations for both modalities by respectively clustering cross-modal data points into classes to enhance intra-cluster correlation. TNN-C-CCA~\cite{zeng2020deep} and VAE-CCA~\cite{zhang2022} improve C-CCA to enhance the correlation of C-CCA by triplet loss and VAE methods. ACMR~\cite{wang2017adversarial}, and AGAH~\cite{GuGGLXW19} utilize adversarial learning to enhance cross-modal representation discrimination. DSCMR~\cite{zhen2019deep} and EICS~\cite{zeng2023learning} achieve discriminative features by introducing representation and label spaces. CLIP~\cite{pmlr_v139_radford21a} and VideoadViser~\cite{wang2023videoadviser} learn transferable visual models from natural language supervision by using a simple pre-training task. BiC-Net~\cite{hou2021bicnet} and MSNSCA \cite{zhang2023multi} utilize transformer architecture to effectively bridge dual modalities. DCIL~\cite{zheng2020dual} introduces an instance loss to improve rank loss by finding appropriate triplets. Conversely, the CCTL~\cite{zeng2022complete} considers all the triplets across two modalities. TLCA~\cite{zeng2023two} employs a progressive training approach, prioritizing easy-to-hard triplet learning over a single-pass model training strategy.

The AV-CMR performances of our method and comparison approaches on two audio-visual datasets are shown in Table~\ref{table:comparison}. From the results, we can observe that our proposed AADML model can achieve the best results on both datasets all over the MAP metrics, with gains of 3.50\%, 2.6\%, and 3.0\% on VEGAS, 48.0\%, 43.2\%, and 45.6\% on AVE in terms of A2V, V2A, and Average, which indicate the effectiveness of our proposed model. Fig.~\ref{fig:PSC} illustrates the variation in average precision across scope@K, where $K$ ranges from 10 to 1000. The curve demonstrates the superior performance of our method across all ranges. This consistency aligns with the MAP comparison presented in Table~\ref{table:comparison} when employing audio and visual as queries.

\subsection{Ablation Study} 
\begin{figure}[t]
    \centering
    \includegraphics[width=8cm]{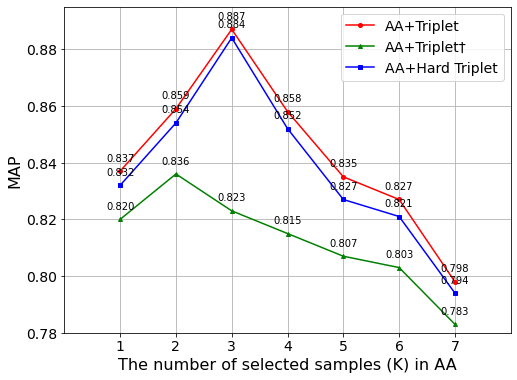}
    \caption{MAP trends on AVE dataset: AA proxy combined with three distinct triplet methods, varying with sample selections (1 to 7) in AA.}
    \label{fig:number}%
\end{figure}

\subsubsection{Impact of the number of selected samples on AA} 
The number of selected samples in AA serves as the sole hyper-parameter in our proposed method. Fig.~\ref{fig:number} presents an experimental analysis of the effect of varying the number of selected samples on AA with three different triplet selection strategies. This exploration aims to prove the significance of this hyperparameter.

For AA combined with either $\textit{triplet loss}$ or $\textit{hard triplet loss}$, the system's performance displays a consistent upward track until approximately 3 samples are selected. Beyond this point, a decrement in performance is observed. In contrast, in the scenario of AA combined with $\textit{triplet$\dag$ loss}$, the performance attains its peak when 2 samples are chosen, following which a decline sets in. These trends result from the interaction between selected samples and AA methodology. In conclusion, the most optimal outcome achieved by AADML occurs through the combination of AA and Triplet techniques, particularly when the AA approach involves the selection of three samples.

\begin{figure}[t]
    \centering
    \includegraphics[width=1.0\columnwidth]{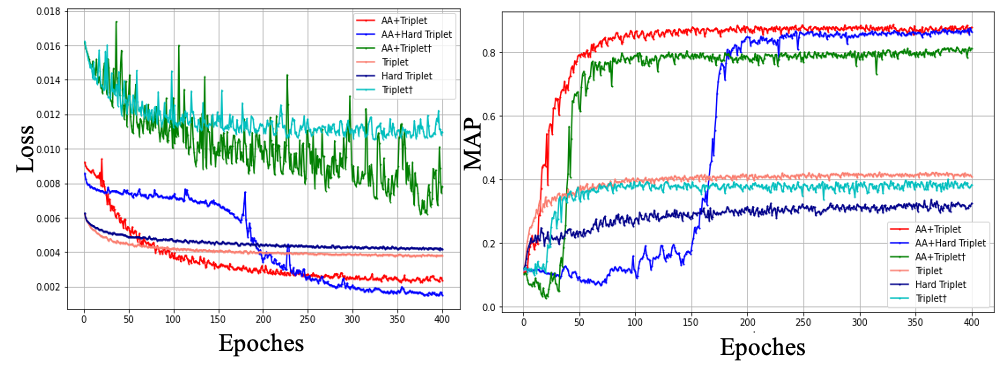}
    \caption{Loss value and MAP performance of training and test set from AVE dataset. Comparative analysis of triplet loss variants: exploring Triplet\dag, Triplet, and Hard Triplet losses with and without AA.}
    \label{fig:epochmap}%
\end{figure}

\begin{table}[t]
\centering

\begin{NiceTabular}{l|ccc}
\toprule
\Block{2-1}{\textbf{Models}} &  \Block{1-3}{\textbf{VEGAS Dataset}} \\ \cline{2-4}
                       & A$\rightarrow$V & V$\rightarrow$A  & Avg. \\ \hline
    Triplet 
    & 0.750 &0.718 &0.734 \\
    Triplet\dag 
    &0.766 &0.765 &0.766 \\
    Hard Triplet 
    & 0.761 & 0.747 & 0.754 \\
    DAML (Triplet) 
    & 0.792 & 0.783 & 0.788\\
    HDML (Triplet) 
    & 0.779 & 0.784 & 0.781\\
    AA+Triplet\dag 
    & 0.867 & 0.856 & 0.862\\
    AA+Hard Triplet 
    &0.865 &0.851 &0.858\\
    \textbf{AA+Triplet (ours)}
    &\textbf{0.901} &\textbf{0.890} &\textbf{0.896}\\ \hline
    Contrastive 
    & 0.701 & 0.702 & 0.704\\
    DSML (Contrastive)
     & 0.745 & 0.736 & 0.41\\
     SupCon (Contrastive)
     & 0.732 & 0.727 & 0.730\\
     CrossCLR (Contrastive)
     & 0.683 & 0.675 & 0.679\\
    AA+Contrastive 
    & 0.783 & 0.807 & 0.795\\ \hline
    N-pair 
    & 0.563 & 0.548 & 0.555\\
    DSML (N-pair)
     & 0.606 & 0.578 & 0.592\\
     SupCon (N-pair)
     & 0.621 & 0.589 & 0.605\\
    AA+N-pair 
    & 0.647 & 0.612 & 0.629\\ \hline
    Angular 
    & 0.581 & 0.533 & 0.557\\
    AA+Angular 
    & 0.613 & 0.549 & 0.581\\ \hline
    Hinge 
    & 0.704 & 0.698 & 0.702\\
    AA+Hinge 
    & 0.761 & 0.764 & 0.762 \\ \hline
    DSL 
    & 0.694 & 0.692 & 0.693\\
    AA+DSL 
    & 0.758 & 0.752 & 0.755\\
    \bottomrule
\end{NiceTabular}
\caption{VEGAS Dataset: The impact of anchor-aware proxy on different metric learning methods. the best MAP scores are presented in bold.}
\label{table:aa_vegas}
\end{table}
\subsubsection{Impact of Triplet Selection Strategy}
The triplet selection strategy directly impacts model performance by determining the quality and diversity of training triplets. Well-chosen triplets facilitate effective learning of data distribution and improve the model's ability to discriminate between categories. In this subsection, we explore the influence of different triplet selection strategies on model training and evaluation using the AVE dataset. 

The first strategy involves selecting all possible triplets within a batch, where a triplet consists of an anchor (1st modality), a positive (2nd modality), and a negative (2nd modality), denoted as $Triplet$. The second strategy revolves around choosing a single triplet from each batch, wherein the triplet comprises an anchor (1st modality), a positive (2nd modality), and the hardest negative (2nd modality), designated as $\textit{Hard Triplet}$. The third strategy derived from this work~\cite{zeng2022complete} entails selecting all available triplets within a batch, where a triplet is composed of an anchor (1st or 2nd modality), a positive (1st or 2nd modality), and a negative (1st or 2nd modality), denoted as $Triplet\dag$.

To investigate the impact of the triplet selection strategy with AA, we conduct experiments involving these three types of triplets with and without the integration of AA, as illustrated in Fig.~\ref{fig:epochmap}. It is evident that incorporating AA significantly enhances the performance of the triplets compared to scenarios without AA overall. Remarkably, the AA+Triplet configuration achieves exceptional results during both the training and testing phases. However, the AA+Hard Triplet strategy exhibits suboptimal performance prior to approximately 150 epochs. Interestingly, before this threshold, the performance of AA+Hard Triplet even lags behind that of the baseline without AA. Notably, this phenomenon is accompanied by considerable fluctuations in the model's loss value, possibly due to the varying detection of similar pairs within different batches. The performance begins to exhibit improvement around the epoch before 150, indicating the model's convergence on the training set.

\begin{table}[t]
\centering

\begin{NiceTabular}{l|ccc}
\toprule
\Block{2-1}{\textbf{Models}} &  \Block{1-3}{\textbf{AVE Dataset}} \\ \cline{2-4}
                       & A$\rightarrow$V & V$\rightarrow$A  & Avg. \\ \hline
    Triplet   
    & 0.404 &0.431 &0.418\\ 
    Triplet\dag 
    & 0.333 &0.351 &0.342 \\
    Hard Triplet 
    & 0.259 &0.355 &0.307 \\
    DAML (Triplet) 
    & 0.370  &0.379 &0.375\\
    HDML (Triplet) 
    & 0.339 &0.293 & 0.316\\
    AA+Triplet\dag 
    & 0.839 & 0.833 & 0.836\\
    AA+ Hard Triplet 
    &0.883  &0.885 &0.884 \\
    \textbf{AA+Triplet} (ours) 
    &\textbf{0.890} &\textbf{0.883} &\textbf{0.887}\\ \hline
    Contrastive   
    & 0.371  & 0.375 & 0.373\\
    DSML (Contrastive)
     & 0.482 & 0.463 & 0.473\\
     SupCon (Contrastive)
     & 0.478 & 0.455 & 0.467\\
     CrossCLR (Contrastive)
     & 0.402 & 0.376 & 0.389\\
    AA+Contrastive 
    & 0.798 & 0.854 & 0.826\\ \hline   
    N-pair 
    & 0.321 & 0.335 & 0.328\\ 
    DSML (N-pair)
     & 0.467 & 0.448 & 0.458\\
     SupCon (N-pair)
     & 0.474 & 0.443 & 0.459\\
    AA+N-pair 
    & 0.828 & 0.833 & 0.830\\ \hline 
    Angular   
    & 0.326 & 0.333 & 0.330\\
    AA+Angular   
    & 0.833  & 0.838 & 0.836\\ \hline  
    Hinge 
    &0.358 &0.381 &0.369\\ 
    AA+Hinge 
    & 0.786 & 0.822 &0.804\\ \hline     
    DSL  &0.371 &0.369 &0.370 \\
    AA+DSL 
    & 0.731 & 0.753 & 0.742\\ 
    \bottomrule
\end{NiceTabular}
\caption{AVE Dataset: The impact of anchor-aware proxy on different metric learning methods. the best MAP scores are presented in bold.}
\label{table:aa_ave}
\end{table}


\subsubsection{Impact of AA Proxy on Metric Learning}
In this subsection, we leverage the assessment of the AA Proxy's impact on existing metric learning techniques, seen in Table \ref{table:aa_vegas} and Table \ref{table:aa_ave}. We integrate our AA proxy with existing novel deep metric learning methods including contrastive loss and triplet loss, and compare them against the state-of-the-art advanced methods designed to enhance these deep metric learning techniques. This comparative analysis aims to determine whether the combination of our AA Proxy with deep metric learning techniques can yield improvements in the performance of AV-CMR tasks. 

The experimental results notably demonstrate the substantial potential of our AA proxy in enhancing the capabilities of deep metric learning methods for AV-CMR. Furthermore, our approach achieves a significant performance advantage over DSML~\cite{duan2018deep}, HDML~\cite{zheng2019hardness}, SupCon~\cite{khosla2020supervised}, and CrossCLR~\cite{zolfaghari2021crossclr} methods when combined with deep metric learning techniques.

To highlight the efficacy of AA as a complementary enhancement for established deep metric learning methodologies, we extend the application of AA proxies to other advanced metric learning methods, including N-pair loss~\cite{sohn2016improved}, Angular loss~\cite{sohn2016improved}, Hinge loss~\cite{bailer2017cnn}, and DSL loss~\cite{cheng2021improving}. The results conclusively reveal that employing AA proxies in conjunction with these advanced deep metric learning methods consistently yields superior results compared to scenarios where AA is not applied.

\section{Conclusion} \label{conclusion}


In this work, we address the challenges of metric learning, which seeks to narrow the gap between similar pairs and enhance the separation of dissimilar pairs for the AV-CMR task. While recent methods show promise by selecting impactful data points during training, the limited number of training samples restricts the full exploration of the embedding space, resulting in an incomplete representation of data distributions. To overcome this, we propose an innovative Anchor-aware Deep Metric Learning approach that adeptly navigates the embedding space even with limited data. Our method simultaneously calculates attention-driven dependencies, considering each sample as an anchor alongside its semantically similar samples. This dynamic correlation weighing within the underlying manifold structure yields Anchor-Aware (AA) scores. Leveraging the parallel computation of AA scores, we employ them as proxies to compute relative distances within the metric learning framework. Extensive experiments on benchmark datasets affirm the effectiveness of AADML, outperforming state-of-the-art models. Furthermore, our exploration of combining AA proxies with various metric learning methods underscores the potency of our approach in advancing the field. In the future, we aim to leverage our methodology for cross-modal retrieval across various modalities and to further expand its applicability into the realm of unsupervised learning.

\bibliographystyle{ACM-Reference-Format}
\bibliography{refs}

\end{document}